\documentclass[namedreferences]{kluwer}
\usepackage{graphicx}
\usepackage{txfonts}
\newcommand{\msol}{\mbox{M}_{\odot}}
\newcommand{\teff}{T_\mathrm{eff}}
\newcommand{\logg}{\log\left(g\right)}
\newcommand{\logy}{\log\left(y\right)}
\newcommand{\kel}{\mbox{K}}

\newcommand{\dex}{\mbox{dex}}
\newcommand{\mgi}{Mg\,I}
\newcommand{\hal}{\mbox{H}\alpha}

\def\farcs{\hbox{$.\!\!^{\prime\prime}$}}
\def\apj{ApJ}%
\def\apjs{ApJS}%
\def\aap{A\&A}%
\def\aaps{A\&AS}%
\def\mnras{MNRAS}%
\begin{document}
\begin{article}
\begin{opening}

  \title{Spectroscopic analysis of sdB stars from the ESO Supernova Ia Progenitor
  Survey\thanks{Based on observations collected at the Paranal
  Observatory of the European
  Southern Observatory for program No.~165.H-0588(A) and 167.D-0407(A).}}

  \author{T. \surname{Lisker}}
  \author{U. \surname{Heber}}
  \author{R. \surname{Napiwotzki}}
  \institute{Dr.\,Remeis-Sternwarte, Bamberg, Germany}
  \author{N. \surname{Christlieb}}
  \author{D. \surname{Reimers}}
  \institute{Hamburger Sternwarte, Hamburg, Germany}
  \author{D. \surname{Homeier}}
  \institute{University of Georgia, Athens, USA}

\begin{abstract}
  We report on the analysis of high-resolution optical spectra for 77
  subdwarf B (sdB) stars from the ESO Supernova Ia Progenitor
  Survey. Effective temperature, surface gravity, and photospheric
  helium abundance are determined simultaneously by spectral line
  profile fitting of hydrogen and helium lines, and are found to be in
  agreement with previous studies of sdB stars. 24 objects show spectral signs
  of a cool companion, being either companion absorption lines or a
  flux contribution at $\hal$. Five stars with relatively high
  luminosity show peculiar $\hal$ profiles, possibly indicating
  stellar winds. Our results are
  compared to recent theoretical simulations by Han et al.\ (2003) for the distribution in
  effective temperature and surface gravity, and are found to agree
  very well with these calculations. Finally we present a binary
  system consisting of two helium-rich hot subdwarfs.
\end{abstract}

\end{opening}


  \section{Introduction \label{sec_int}}

  In the past years, subdwarf B (sdB) stars were subject to many
  observational and also theoretical studies, raising lots of new
  questions in attempting to answer the old ones. How are these
  stars formed? The relative importance of single-star formation as
  well as binary evolution still needs to be determined from
  observations. Maxted et al.\ (2001) showed that many sdB stars reside in close
  binaries, indicating a former mass transfer phase to account for the
  thin envelope of the star. Han et al.\ (2003) examined various formation
  channels for simulating sdB formation in this
  context, combining them to yield the observed sdB
  population. Theoretical studies of this kind need to be compared to
  observational data of high quality to evaluate the simulation
  results and judge our current understanding of sdB formation. With
  our sdB sample from the ESO Supernova Ia Progenitor
  Survey (SPY, Napiwotzki et al.~2001), we present a homogeneous, high
  quality dataset of higher resolution and larger wavelength coverage
  than previous studies of comparable size. Its analysis should
  therefore bring new insight into the physics of sdB stars.


  \section{Observation, reduction, and spectral classification\label{sec_obs}}

  Spectra were obtained at ESO VLT with the UV-Visual Echelle Spectrograph
  (UVES) at UT2 (Kueyen). A slit width of
  $2\farcs1$ was used, resulting in a resolving power of $18\,500$
  ($0.36\,\AA$ at $\hal$) or better. The wavelength coverage is
  $3300-6650\,\AA$, with gaps at about $4500-4600\,\AA$ and
  $5600-5700\,\AA$. For most of the stars, two spectra in different
  nights were taken. They
  were then reduced with a procedure developed by C.~Karl using the ESO MIDAS
  software package and parts of the UVES automatic reduction
  pipeline. The reduced spectra
  were convolved with a Gaussian of $1.0\,\AA$ FWHM and
  rebinned to $0.4\,\AA$ pixel size.
  
  After a first selection of potential sdB spectra from SPY by visual
  inspection, a rough line profile fit was performed, and objects with
  helium-dominated atmospheres were excluded. Following the common
  assumption that sdB and sdOB stars belong to the same evolutionary
  type, all selected stars are named sdB stars, being 77 objects with
  $20\,000\,\kel < \teff < 38\,000\,\kel$, $4.8 < \logg < 6.0$ and
  $-4.0 \le \logy < -0.8$ (photospheric helium abundance $y
  =N_\mathrm{He}/N_\mathrm{H}$). 51 program stars are objects
  from the Hamburg/ESO survey (Wisotzki et al.~1996), 9 from the Hamburg Quasar
  Survey (Hagen et al.~1995), 14 from McCook \& Sion (1999), and 3 from other sources.

  In many spectra, signs of a cool
  companion can be seen, e.g.~a broad Ca\,K line ($3933\,\AA$) as
  opposed to the frequently occuring narrow interstellar Ca\,K
  line. The \mgi~triplet at $5167\,\AA, 5173\,\AA$, and $5184\,\AA$
  turned out to be a good indicator for cool companions, because it
  reveals even the presence of companions with a very low contribution
  to the total flux. In the spectra of five stars, though, only a
  rise in the continuum level at $\hal$ pointed towards a composite
  object. Altogether, 24 out of 77 sdB
  stars show signs of a cool companion. These objects are
  excluded from the following analysis, because their determined
  parameters cannot be considered reliable.


  \section{Results from spectral analysis \label{sec_res}}

  The determination of $\teff$, $\logg$, and $\logy$ was done by fitting
  simultaneously each hydrogen
  and helium line (except $\hal$) to synthetic model spectra, using a procedure
  by R.~Napiwotzki (Napiwotzki et al.~1999) based on Saffer et al.\ (1994). See
  Edelmann et al.\ (2003) for a more detailed description. 
  The final results were calculated as the mean value of the fit results
  from individual exposures of each star. For six objects, only one
  useful exposure was available.

  The statistical 1-$\sigma$-errors yielded by the fit procedure are
  typically lower than $100\,\kel$, $0.02\,\dex$, and $0.04\,\dex$ for
  $\teff$, $\logg$, and $\logy$, respectively. However, since the
  differences in the fit results of individual exposures of each star
  are mostly larger, we used the distribution of those differences for
  estimating the ``true errors'', leading to $\Delta\,\teff =
  360\,\kel$, $\Delta\,\logg = 0.048\,\dex$, $\Delta\,\logy =
  0.045\,\dex$.

  \subsection{Atmospheric Parameters \label{sec_atmo}}

  \begin{figure}
    \centerline{\includegraphics[width=17pc]{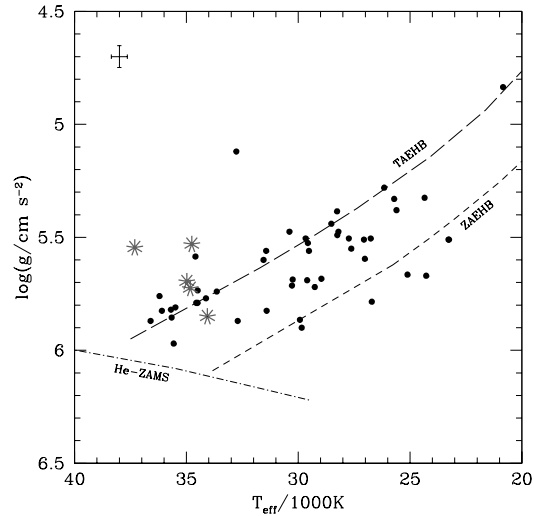}}
    \caption{Distribution of our sdB sample in the
      $\teff$-$\logg$-plane. Grey star symbols denote objects with
      peculiar $\hal$ profiles (see Sect.~\ref{sec_hal}).}
    \label{myehb}
  \end{figure}

  \begin{figure}
    \centerline{\includegraphics[width=17pc]{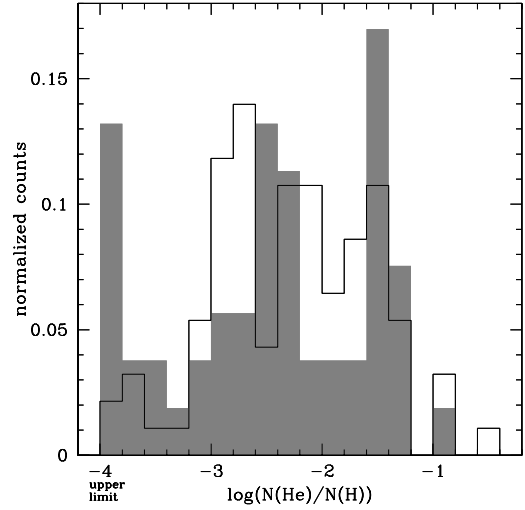}}
    \caption{\emph{Shaded histogram:} Helium abundance distribution
    of our sdB stars. \emph{Open histogram:} Distribution from
    Edelmann et al.\ (2003).}
    \label{hehist}
  \end{figure}

  Figure \ref{myehb} shows the position of our sdB stars in the $\teff$-$\logg$-plane,
  along with the Zero-Age EHB and Terminal-Age EHB for solar
  metallicity (ZAEHB and TAEHB, Dorman et al.~1993)
  and the Zero-Age Main Sequence for pure helium stars (He-ZAMS,
  Paczy\'nski 1971). Up to
  effective temperatures of $\teff \approx 33\,000\,\kel$, most of our stars lie
  on the so-called EHB strip (defined by ZAEHB and TAEHB), while for
  higher $\teff$, almost all objects lie near the TAEHB or above
  it. This behaviour is reproduced in current theoretical
  calculations, as discussed in Sect.~\ref{sec_han}.

  Our distribution of helium abundances (Fig.~\ref{hehist}) shows a wide spread
  from $\logy < -4$ to slightly supersolar
  helium abundance. Compared with the values from
  Edelmann et al.\ (2003) shown in Fig.~\ref{hehist}, there is overall agreement, but we find a higher fraction
  of stars with very low helium abundances, lacking an explanation. Theoretical
  calculations of the photospheric helium content, which include mass
  loss rates of $\dot{M} =10^{-14}$ to $10^{-13}\,\msol/\mbox{yr}$, are
  able to explain
  values of $-4 < \logy < -2$ (Fontaine \& Chayer 1997; Unglaub \&
  Bues 2001). In our data, 18 out
  of 53 stars (one third) have $\logy > -2$, clearly pointing towards
  higher mass loss rates in many cases or alternative (additional)
  physical mechanisms.

  \subsection{Peculiar Halpha profiles \label{sec_hal}}

  \begin{figure}
    \centerline{%
      \begin{tabular}{c@{\hspace{2pc}}c}
	\includegraphics[width=12pc]{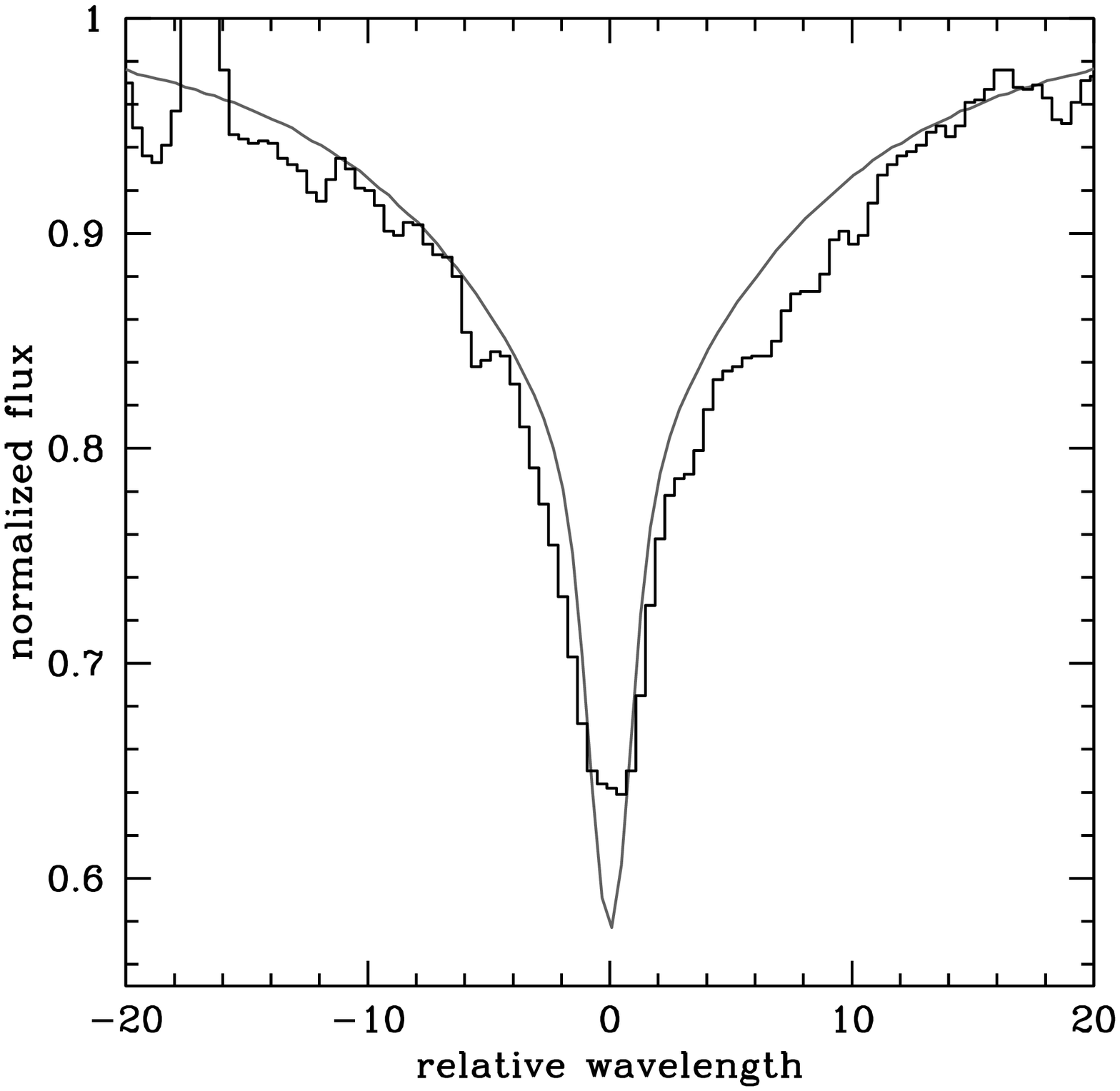} &
	\includegraphics[width=12pc]{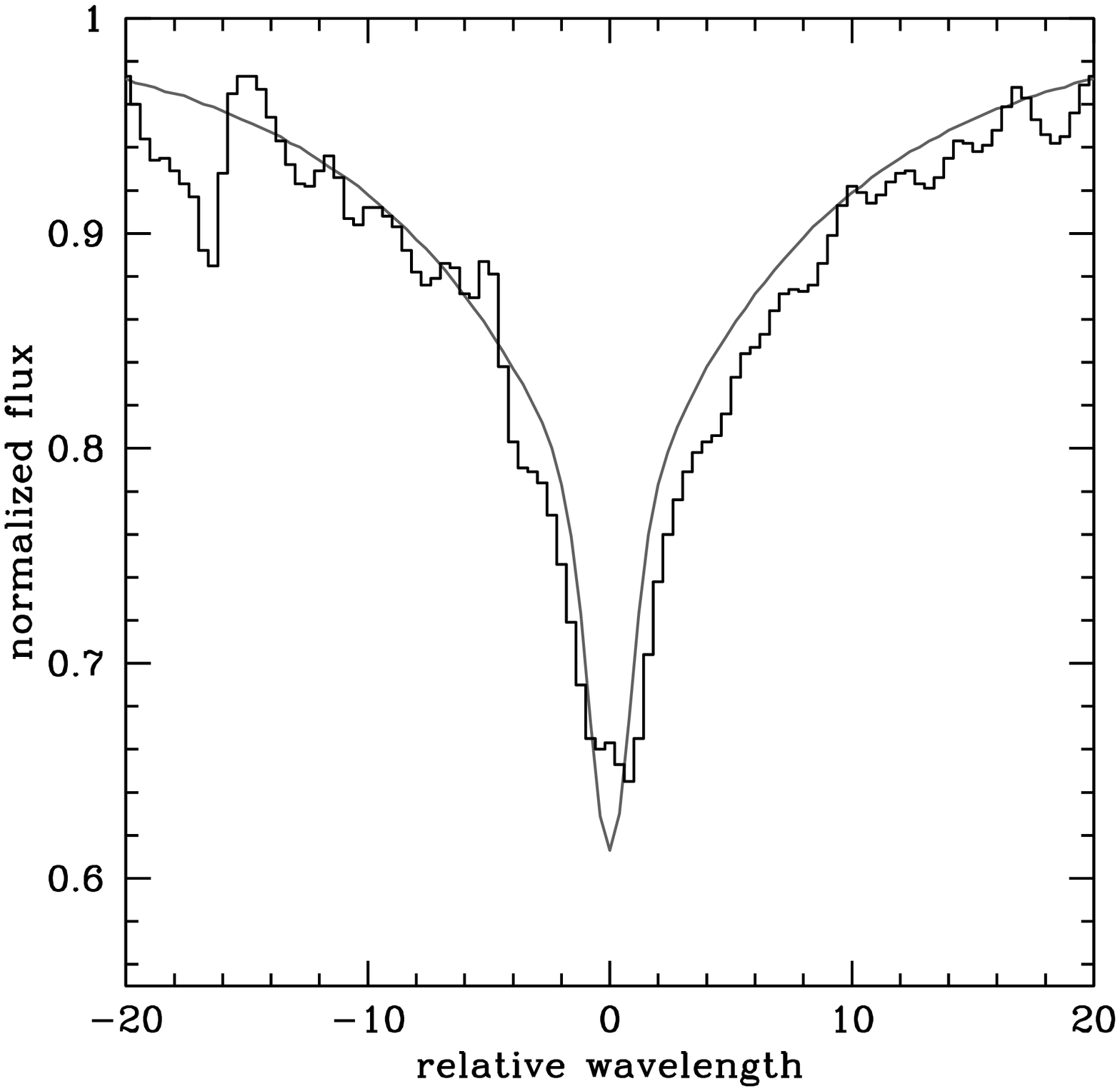} \\
      \end{tabular}}
      \caption{Two spectra with peculiar $\hal$ profiles, the left
    with a flat line core, the right with a faint emission. Observed
    spectra are shown as histograms, model spectra as solid grey lines.}
      \label{halpha}
  \end{figure}

  Five of our sdB stars show abnormal
  $\hal$ line core profiles (Fig.~\ref{halpha}), having a lower depth than the model
  line core, being broader, and in two cases showing a faint emission. No
  significant variation is observed between the first and second
  exposure of the stars (time intervals range from $3\,\mathrm{d}$ to
  $1\,\mathrm{yr}$), but slight changes cannot be excluded. All objects
  are amongst the hottest program stars with $\teff > 34\,000\,\kel$
  (Fig.~\ref{myehb}), and four lie among the highest luminosity
  objects of the sample (the luminosity of the fifth sdB is
  about average). The helium abundances are in the normal range of
  $-3.5 < \logy < -1.1$.

  Line broadening could possibly be caused by rotation. However,
  convolution of the synthetic spectra with rotational profiles does
  not lead to acceptable fits for any hydrogen or helium line. Even for $\hal$ itself, the
  match is still not satisfying, implying that other physical effects have
  to be the cause.

  There is an obvious similarity to the findings of
  Heber et al.\ (2003), who observed such $\hal$ profiles in
  four high-luminosity sdB stars and suggested
  stellar winds to be the cause. Their profiles
  also show broadening around the core region, with the core being
  flattened in one case and containing emission in the
  other cases. As shown by Vink (2003, these proceedings), it is possible to
  reproduce the emission when calculating winds of $\dot{M}
  =10^{-11.5}\,\msol/\mathrm{yr}$. The broadening, however, is not
  accounted for, and still remains to
  be explained.

  \subsection{Comparison with other samples \label{sec_comp}}

  \begin{figure}
    \centerline{\includegraphics[width=17pc]{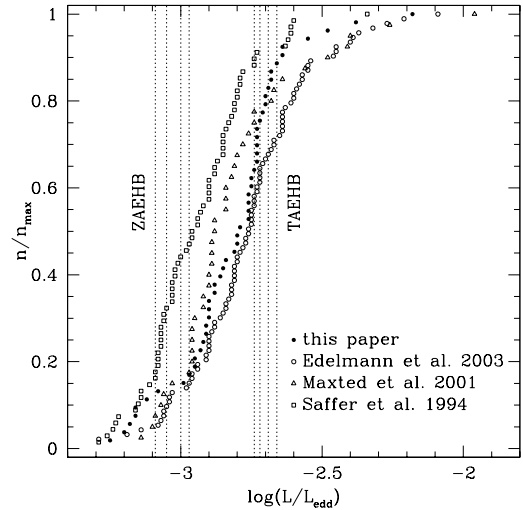}}
    \caption{Luminosity in Eddington luminosities versus cumulative
    normalized counts from this work as well as from three previous
    studies. ZAEHB and TAEHB are for $[\mathrm{Fe}/\mathrm{H}] = 0.00,
    -0.47, -1.48, -2.26$ from left
    to right (Dorman et al.~1993).}
    \label{cumlum}
  \end{figure}

  SdB samples of comparable size were presented in the last decade
  by Saffer et al.\ (1994), Maxted et al.\ (2001), and Edelmann et al.\ (2003). They can best be
  compared to our sample by using their respective cumulative
  luminosity distribution, shown in Fig.~\ref{cumlum}.
  It can be seen that the SPY data agree well
  with the observations of Maxted et al.\ (2001) and Edelmann et al.\ (2003), whereas
  there is a larger average offset to the data of Saffer et al.\ (1994) (which has
  already been discussed in Edelmann et al.~2003). Different resolution,
  signal-to-noise ratio, homogeneity, and models used for line profile fitting can
  probably explain these moderate differences between the samples.

  \section{Observation versus Theory \label{sec_han}}

  \begin{figure}
    \centerline{\includegraphics[width=17pc]{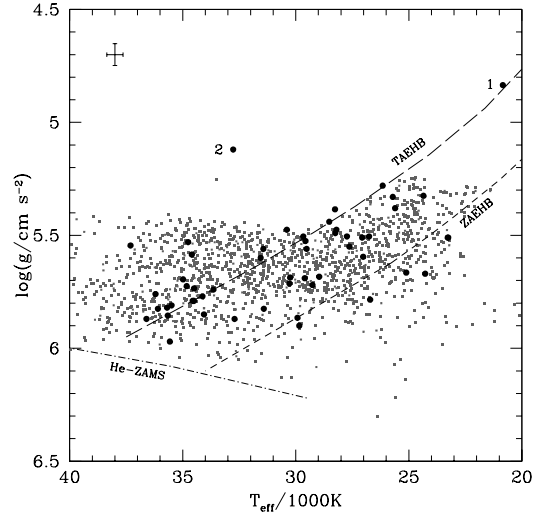}}
    \caption{$\teff$-$\logg$-distribution of our sdB sample (filled circles) along
      with the simulated stars from the best-fit model of Han et
      al.\ (2003, grey dots).}
    \label{myhan}
  \end{figure}

  In order to draw conclusions about potential sdB formation scenarios, we now compare our data with the
  simulations of Han et al.\ (2003), which aim to determine the relative importance of
  different sdB binary formation channels (stable
  Roche-lobe overflow channel, common envelope ejection channel,
  helium white dwarf merger channel). Several poorly known
  physical parameters, e.g.~common envelope ejection efficiency, are
  varied in order to produce twelve simulation sets with different
  parameter configurations. Out of them, Han et al.\ (2003) determined a best-fit model
  by comparing periods and minimum companion masses for binary sdB stars with
  observations by Maxted et al.\ (2001) and Morales-Rueda et al.\ (2003). A comparison with
  our $\teff$-$\logg$-data is interesting because of the
  homogeneity and high resolution of our sdB spectra, surpassing previous
  observational samples.

  Before comparing theory and observation, one must consider
  observational selection effects, in the present case mainly due to
  main sequence (MS) companions, which either outshine the sdB and thus
  prevent its detection, or make its analysis
  unreliable. Han et al.\ (2003) apply this effect by excluding all systems
  with a MS companion having $\teff > 4000\,\kel$, which is matched by
  our exclusion of all composite spectra and therefore allows us to
  compare our data with the theoretical predictions.

  Figure \ref{myhan} shows our data along with the simulated sdB stars
  from the best-fit model of Han et al.\ (2003). The simulations match well
  the shape of our sdB distribution as well as the
  $\teff$-$\logg$-area covered. One of our objects is a potential
  blue Horizontal Branch star (labeled ``1'' in Fig.~\ref{myhan}) and
  is therefore not subject to a comparison with calculations for EHB
  stars. The object labeled ``2'' lies in a region where there is
  almost no simulated star. However, this object may be in a fast post-EHB
  phase of evolution (see also Fig.~\ref{myehb}). As long as
  theory does not predict the formation of such an object to be
  impossible, it cannot be considered a disagreement with the
  simulations to observe one such star.

  The positions of our stars at or above the TAEHB at higher
  temperatures are reproduced in the simulations, partially reflecting
  their prediction of masses in the range $0.4 <
  M_\mathrm{sdB} < 0.6\,\msol$, as opposed to $0.46 <
  M_\mathrm{sdB} < 0.50\,\msol$ from single star evolutionary
  calculations (which define the ZAEHB and TAEHB shown in
  Fig.~\ref{myhan}) . Note, however, that the latter predict some
  $10\,\%$ of the objects to be in the post-EHB stage (Dorman et al.~1993), which could
  probably explain the observed stars at high temperatures and
  luminosities above the TAEHB as well.


  \section{A subdwarf\,+\,subdwarf binary \label{sec_2x}}

  \begin{figure}
    \centerline{%
      \begin{tabular}{c@{\hspace{2pc}}c}
	\includegraphics[width=12pc]{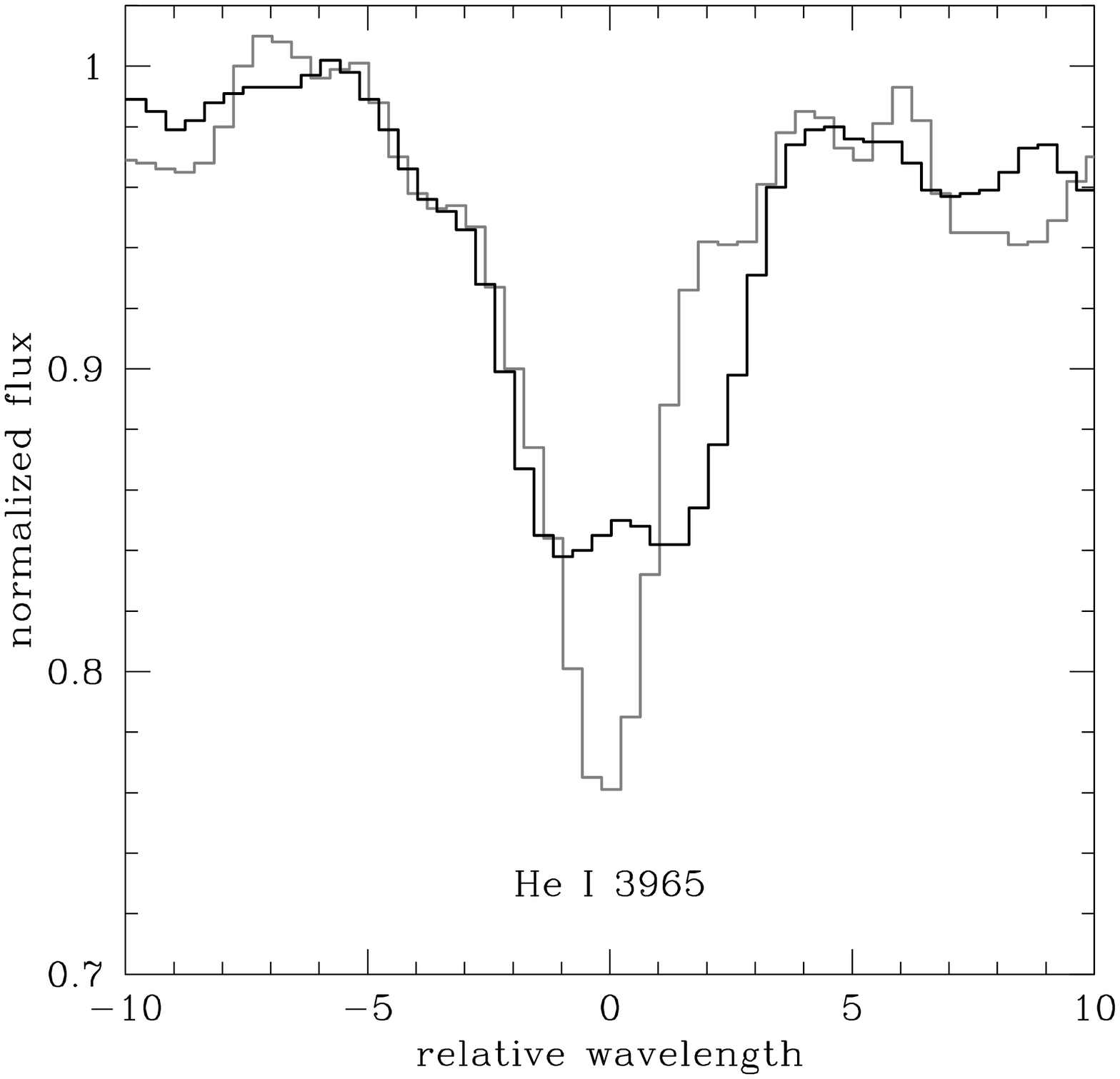} &
	\includegraphics[width=12pc]{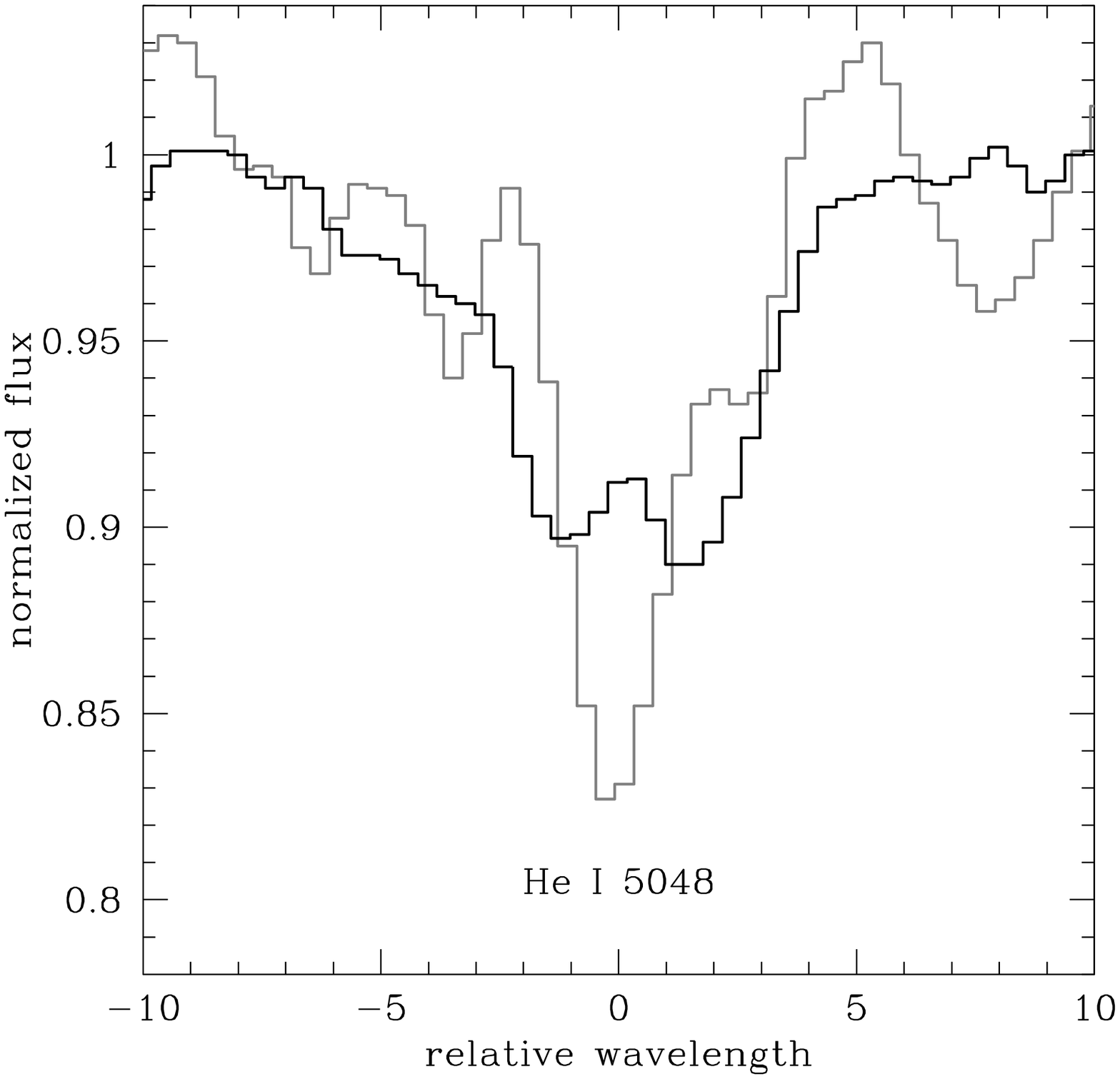} \\
      \end{tabular}}
      \caption{Comparison of first exposure (grey histogram) and
      second exposure (black histogram) of HE
      0301$-$3039. Both spectra were brought to the same flux
      level. Barycentric corrections have been applied, and the given
      wavelength is relative to the apparent line center.}
      \label{2x}
  \end{figure}

  Figure \ref{2x} shows parts of two spectra
  of the helium-rich hot subdwarf HE 0301$-$3039, the time
  interval being $10.5$ months. The lines split into two
  apparently equal components, being obviously deeper when combined
  than when separated, the radial velocity difference being about
  $160\,\mathrm{km/s}$ for all lines. Hence HE 0301$-$3039 is
  a double-lined binary system consisting of two very similar hot
  subdwarfs. To our knowledge this is the first such system
  discovered. Further observations are scheduled to measure its radial
  velocity curve and system parameters.
  

  \section{Conclusions}

  We have reported on the spectral analysis of 77 sdB stars
  from SPY, as well as on the detection of the first known binary consisting of two hot
  subdwarfs. 24 sdB stars show signs of a cool companion. Of the 53
  non-composite objects, five stars with relatively high luminosity
  show peculiar $\hal$ profiles, possibly indicating stellar winds. Our data are found to be in good
  agreement with previous studies of sdB stars. The best-fit model of the binary population
  synthesis calculations by Han et al.\ (2003) reproduces the observed sdB
  distribution very well. Note, however, that at this point no statement can be made as
  to the relative importance of single-star formation channels. A detailed comparison of our observations with
  all simulation sets of Han et al.\ (2003) will be presented in a
  forthcoming paper.


  \begin{acknowledgements}
    We would like to thank Zhanwen Han for providing us with his simulation
    results, and Jorick S.~Vink for valuable information on stellar
    winds.
  \end{acknowledgements}

\section*{References}
\begin{small}
Dorman, B., Rood, R.~T., \& O'Connell, R.~W. 1993, \apj, 419, 596\smallskip

\noindent Edelmann, H., Heber, U., Hagen, H.-J., et~al. 2003, \aap, 400, 939\smallskip

\noindent Fontaine, G. \& Chayer, P. 1997, in The Third Conference on Faint Blue
  Stars, 169\smallskip

\noindent Hagen, H.-J., Groote, D., Engels, D., \& Reimers, D. 1995, \aaps, 111,
  195\smallskip

\noindent Han, Z., Podsiadlowski, P., Maxted, P.~F.~L., \& Marsh, T.~R. 2003,
  \mnras, 341, \indent 669\smallskip

\noindent Heber, U., Maxted, P.~F.~L., Marsh, T.~R., Knigge, C., \& Drew, J.~E.
  2003, in ASP \indent Conf. Ser. 288: Stellar Atmosphere Modeling, 251\smallskip

\noindent Maxted, P.~F.~L., Heber, U., Marsh, T.~R., \& North, R.~C. 2001,
  \mnras, 326, 1391\smallskip

\noindent McCook, G.~P. \& Sion, E.~M. 1999, \apjs, 121, 1\smallskip

\noindent Morales-Rueda, L., Maxted, P.~F.~L., Marsh, T.~R., North, R.~C., \&
  Heber, U. 2003, \indent \mnras, 338, 752\smallskip

\noindent Napiwotzki, R., Christlieb, N., Drechsel, H., et~al. 2001,
  Astron.~Nachr., 322, 411\smallskip

\noindent Napiwotzki, R., Green, P.~J., \& Saffer, R.~A. 1999, \apj, 517, 399\smallskip

\noindent Paczy\'nski, B. 1971, Acta Astronomica, 21, 1\smallskip

\noindent Saffer, R.~A., Bergeron, P., Koester, D., \& Liebert, J. 1994, \apj,
  432, 351\smallskip

\noindent Unglaub, K. \& Bues, I. 2001, \aap, 374, 570\smallskip

\noindent Wisotzki, L., Koehler, T., Groote, D., \& Reimers, D. 1996, \aaps, 115,
  227
\end{small}

\end{article}
\end{document}